\documentclass[useAMS,usenatbib,letters]{mn2e}

\usepackage{graphicx}
\usepackage{color}
\definecolor{orange}{rgb}{1,0.5,0}
\usepackage[colorlinks=True,citecolor=orange,urlcolor=blue,linkcolor=red]{hyperref}

\def\apj{ApJ}%
\def\apjl{ApJ}%
\def\aap{A\&A}%
\def\mnras{MNRAS}%
\begin{document}

\title[MHD Shocks and Sub-critical cores]{The  Responses of Magnetically  Sub-Critical Cores  to Shocks.}
\author[B.          Vaidya,          T.         W.          Hartquist,
S. A. E. G. Falle]{B. Vaidya$^{1}$\thanks{E-mail: B.Vaidya@leeds.ac.uk
(BV)}, T. W. Hartquist$^{1}$, S.  A. E. G. Falle$^{2}$\\
$^{1}$School of Physics and Astronomy, University of Leeds, Leeds LS2
9JT, UK\\
$^{2}$Department of  Applied Mathematics, University  of Leeds, Leeds
LS2 9JT, UK}

\date\today

%\date{Accepted yyyy m dd. Received yyyy m dd; in original yyyy m dd}
\pagerange{\pageref{firstpage}--\pageref{lastpage}} \pubyear{yyyy}

\maketitle

\label{firstpage}

%----------------------------------

\begin{abstract}
{An   ideal  magnetohydrodynamics  (MHD)   code  with   adaptive  mesh
refinement (AMR) was used to investigate the interactions of fast-mode
shocks  with  self-gravitating,  isothermal  cores  with  mass-to-flux
ratios  that  are  somewhat  below  the  minimum  value  required  for
gravitational  collapse.   We   find  that  shock  focussing  produces
colliding  flows  along  the  field  lines  that  generate  very  high
densities, even for relatively weak shocks.  Self-gravity plays only a
minor role in determining the  highest density that is reached, but it
does  play a  role  in  the subsequent  evolution.   The densities  at
comparable  times differ by  a factor  of a  few for  shocks initially
propagating perpendicularly or obliquely  to the magnetic field in the
ambient medium.}
\end{abstract}


%------------------------------------------------------------------
\section{Introduction}
The  importance  of  shock-cloud  interactions for  feedback  in  star
formation
%, the  survival of structure in the  interstellar medium, and
%the regulation of the global properties of the interstellar medium
has  motivated a number  of groups  to perform  3D MHD  simulations of
shocks  interacting   with  single  clouds  \citep{Gregori:2000p10754,
Shin:2008p9750,           Kwak:2009p10827,           VanLoo:2010p9828,
Shelton:2012p10934}.  None of  these simulations included self-gravity
and  only  those of  \citet{VanLoo:2010p9828}  and  some  of those  of
\citet{Shelton:2012p10934}  included radiative  cooling.  Like  the two
dimensional       simulations      of      \citet{Fragile:2005p10978},
\citet{Lim:2005p5897},          and          \citet{VanLoo:2007p9421},
\citet{VanLoo:2010p9828} considered  the production of  cooler regions
by  including   the  thermal  instability  of   warm  phase  material.
\citet{Shelton:2012p10934}  focussed   on  X-ray  emission   and  cloud
destruction rather than the cloud internal structure.

This  paper concerns the  effects of  shocks, which  are likely  to be
driven by the  outflows of recently born stars,  on cores in molecular
clouds.   A  core  is  assumed  to  be  in  an  isothermal  magnetized
equilibrium     state,     such     as     those     considered     by
\cite{Mouschovias:1976p8743,     Mouschovias:1976p8742}     and     is
magnetically  sub-critical i.e.   its mass-to-flux  ratio  is somewhat
below the critical  value for the core to  collapse under gravity. The
analysis of  infrared polarization maps  of some molecular  clouds has
shown  that  they  contain   pc  scale  cores  that  are  magnetically
sub-critical \citep{Chapman:2011p11866, Marchwinski:2012p11767}.  We shall see that,
although  even quite  weak  shocks  can produce  a  large increase  in
density,  this  does  not  lead  to gravitational  collapse  in  ideal
MHD.  However, self-gravity does  retard the  subsequent re-expansion.
We intend to include ambipolar diffusion and the Hall effect in future
work \citep[e.g.,][]{Ashmore:2010p7041}.

\section{Numerical Technique}

The calculations  were performed  with the hierarchical  adaptive mesh
refinement (AMR) code, MG \cite{Falle:2012p12513}. This solves the equations
of ideal  MHD using a second  order upwind scheme with  the linear MHD
Riemann solver  described in \cite{Falle:1998p4496}  combined with the
divergence cleaning technique described in \cite{Dedner:2002p9747}.  A
hierarchy of $N$ grids levels,  $G_0 \cdots G_{N-1}$, is used, and the
mesh spacing for  $G_n$ is $\Delta x/2^{n}$, where $\Delta  x $ is the
cell size  for the coarsest level,  $G_0$.  $G_0$ and  $G_1$ cover the
entire domain,  but finer grids  need not do  so.  Refinement is  on a
cell-by-cell basis and  is controlled by error estimates  based on the
difference between  solutions on different grids  i.e.  the difference
between the  solutions on $G_{n-1}$ and $G_n$  determine refinement to
$G_{n+1}$.   Self-gravity  is  computed  using  a  full  approximation
multigrid to solve the Poisson equation.

\section{Initial Conditions}
\label{sec:InitCond}

The initial core has  density $\rho_i$,  sound speed
$c_c$ and radius $R_i$ and is embedded in a warmer uniform medium with
sound speed  $c_e = 4  c_c$ and pressure  $0.9 \rho_i c_c^2$.  This is
implemented by defining an advected  scalar, $\alpha$, that is unity in
the cloud and zero in the  surroundings. The sound speed, $c$, is then
given by  $c^2 = \alpha  c_c^2 + (1  - \alpha) c_e^2$. This  scalar is
also used to  turn off gravity in the external  medium.  Both the core
and its  surroundings are  threaded by a  uniform magnetic  field with
magnitude $B_0$.   We use dimensionless  units in which $\rho_i  = 1$,
$c_c =  1.0$ and the gravitation  constant $G = 1.0$.   In these units
the core has $R_i = 2.5/\surd{4  \pi} = 0.705$ and a free-fall time of
$(3  \pi/32)^{0.5}$.   For the  adopted  units,  the initial  magnetic
pressure is ${{B_0}^2}/2$ (note that  we have suppressed factors of $4
\pi$ in the equations). 

This  initial  core  is  not   in  equilibrium,  but  evolves  into  an
equilibrium state provided the  mass-to-flux ratio is below a critical
value. This equilibrium  state, which is produced by  the  collapse of  a
uniform, non-rotating,  isothermal, spherical core,  is the same
as   one   of   those   specified   by   \citet{Mouschovias:1976p8743,
Mouschovias:1976p8742}.
For a zero temperature core, the critical value of 
mass to flux ratio is given by \citep{Mouschovias:1976p9250}

\begin{equation}
\frac{M_{\rm            crit}}{\Phi_{\rm            crit}}           =
\frac{0.53}{3\pi}\left(\frac{5}{G}\right)^{\frac{1}{2}}
\label{eq:mtfcr}
\end{equation}

\noindent
Since, in case of ideal MHD, the mass-to-flux ratio does not change, we have

\begin{equation}
\frac{M}{\Phi} = \frac{4 \rho_i R_i}{3 B_0}.
\label{eq:mtf}
\end{equation}

\noindent
We set

\begin{equation}
\lambda = \frac{M}{\Phi} \frac{\Phi_{\rm crit}}{M_{\rm crit}} = 0.707,
\label{eq:lambda}
\end{equation}

\noindent
which gives an initial plasma $\beta$

\begin{equation}
\beta_i = \frac{2 \rho_i c_c^2}{B_0^2} = 0.224.
\label{eq:betai}
\end{equation}
 
The equilibrium  core has an oblate  shape with an  aspect ratio $\sim
0.46$.  The  maximum value  of the density  is $2.08$ and  the maximum
value of $\beta$ is $0.426$.

\begin{figure*}
\includegraphics[width=2.1\columnwidth]{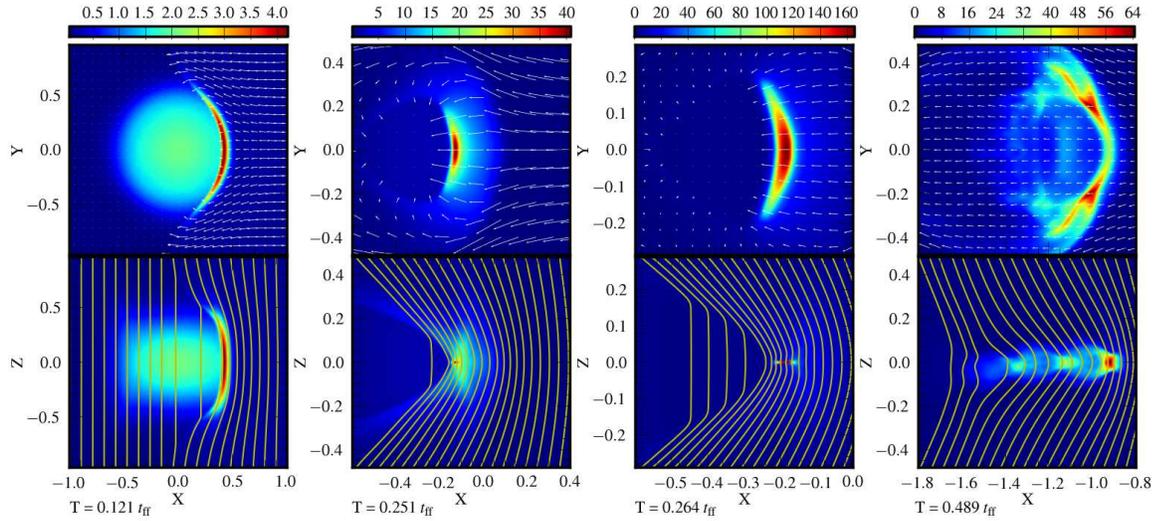}
\caption{Density,  velocity vectors  and  magnetic field  lines for  the
perpendicular shock.  Each of the four columns represents the solution
at the time (expressed in terms of free fall time) indicated below the
panels. The top panels show the z  = 0 plane and the bottom panels the
y =  0 plane.  The  white arrows in  the top panels  indicate velocity
vectors  and  the solid  lines  in  the  bottom panels  represent  the
magnetic  field  lines.  The  colorbar  placed above  the  top  panels
provides the  measure of  core density in  terms of $\rho_i$  for each
column.}
\label{fig:shockevolve}
\end{figure*}

\begin{figure*}
\includegraphics[width=2.1\columnwidth]{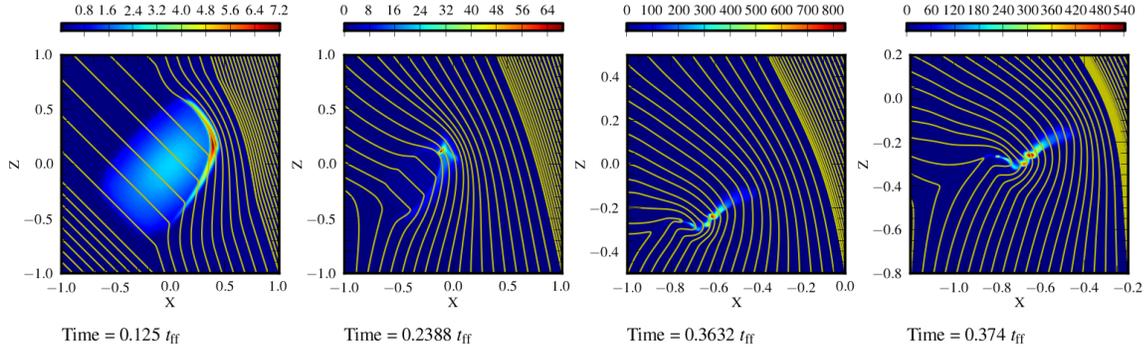}
\caption{As  for \ref{fig:shockevolve}  for the  y =  0 plane  for the
oblique case.}
\label{fig:obshockevolve}
\end{figure*}

All calculations were performed on a three-dimensional cartesian grid,
$-2 \le x \le 2$, $-2 \le y  \le 2$, $-2 \le z \le 2$, with the centre
of  the core  at the  origin.  Initially $6$  grids were  used with  a
resolution  of  $10^3$ on  $G_0$,  which  gives  an effective  maximum
resolution of $320^3$ ($225^3$ for the initial cloud). Note that $G_0$
needs  to  be  coarse in  order  to  ensure  fast convergence  of  the
multigrid Poisson solver.  For the evolution to the equilibrium state,
free flow boundary conditions were imposed on all boundaries.

This resolution is  more than adequate for the  equilibrium state, but
as we shall see, is not  sufficient to resolve the high density region
that is produced by the  shock interaction.  However, the code has the
ability  to change  the  number of  levels  during the  course of  the
calculation, so that additional levels could be added as necessary.

\section{Shock Interaction}
\label{sec:shocks}

A  fast-mode   shock  was  introduced  onto  a   grid  containing  the
equilibrium core by setting the conditions on the $x = 2$ plane to the
postshock state  for such a shock  in the negative  $x$ direction with
and upstream state corresponding of that of the warm medium.

We  consider  two cases:  perpendicular  ($\theta  = 90^{\circ}$)  and
oblique ($\theta = 45^{\circ}$),  where $\theta$ is the angle between
the  shock  normal  and  the  upstream magnetic  field  far  from  the
core. For the perpendicular shocks, the equilibrium core was generated
from an  initial state with the  magnetic field in  the $z$ direction,
but for the oblique shock it was at $45^\circ$ to the $z$ axis.

We chose  to characterize  the strength of  the shock by  its Alf\'ven
Mach number

\begin{equation}
M_{a} = V_{\rm shock}/V_a,
\end{equation} 

\noindent
where, $V_{\rm shock}$  is the shock speed in  the upstream rest frame
and $V_{a}$ is the the Alfv\'en speed given by

\begin{equation}
V_{a} =  B/\sqrt{\rho}.
\end{equation}
in our equations.
\noindent
This has the advantage that an oblique shock has the same speed as a
perpendicular shock with the same value of $M_a$.
 
Fig.~\ref{fig:shockevolve} shows the density for a perpendicular shock
with $M_a = 2.0$ at four  times, measured from the time that the shock
was introduced.   As can be seen  from the figure,  a filamentary high
density region  is formed, which  is highly flattened parallel  to the
magnetic field.  In order to resolve  this, it was necessary to add an
extra  three grid  levels  as  the calculation  proceeded  to give  an
effective resolution of $2560^3$ ($900^3$ for the initial cloud). Even
so, this  is barely sufficient to  resolve the high  density region in
its most compressed state. Fig.~\ref{fig:obshockevolve} shows that the
oblique shock also generates a dense region.

Careful examination of the results  shows that the dense region is the
result of shock  focussing by the density gradient at  the edge of the
core.   A plane  shock  that  encounters such  a  density gradient  is
refracted until  its direction of propagation becomes  parallel to the
density gradient  (much like water waves  on a sloping  beach). In the
perpendicular  case, this leads  to strong  focussing on  the $z  = 0$
plane where the density contours  have a small radius of curvature. As
in a Munro  jet \citep{Birkhoff}, this would lead  to a large pressure
and hence density  even if there were no magnetic  field, but here the
velocities along the field are of the order of the post-shock Alfv\'en
speed, which is  significantly higher than the gas  sound speed in the
core.   As a  result, convergence  along the  field lines  leads  to a
higher density than in the purely hydrodynamic case.

It is possible to estimate the way in which the maximum density scales
with $M_a$ and the initial  $\beta$ in the core.  For a perpendicular
isothermal shock, the compression (see Appendix in \citealt{Yu:2006p12405}) is

\begin{equation}
\tau  = \frac{1}{2}  [-\beta_0  - 1  +  \surd \{(1  +  \beta_0)^2 +  8
M_a^2\}],
\label{exacttau}
\end{equation}

\noindent
where $\beta_0$ is the upstream $\beta$. The post-shock total pressure
is then

\begin{equation}
p = \frac{B_0^2}{2} (\tau^2 + \tau \beta_0),
\label{exactp}
\end{equation}

\noindent
where $B_0$ is the upstream magnetic field.

For the  incident shock in the  low density medium we  have $\beta_0 =
\beta_i$ ($= 0.224$ in our case) and we can ignore the gas pressure to
get

\begin{equation}
\tau_i \simeq \surd 2 M_a
\label{approxtau}
\end{equation}

\noindent
if  we  also ignore  $(1  + \beta_0)^2$  compared  with  $8 M_a^2$.  The
post-shock pressure is then given by

\noindent
\begin{equation}
p_i \simeq \tau_i^2 \frac{B_i^2}{2} \simeq M_a^2 B_i^2,  
\label{approxp}
\end{equation}

\noindent
where $B_i$ is the initial magnetic field.

We can get  a lower limit on the density  behind the shock propagating
into the  core by  assuming a perpendicular  shock with post-shock
pressure $p_i$.  This is clearly  a lower limit since it ignores shock
convergence, the post-shock pressure is  greater than $p_i$ due to the
reflected  shock  and the  compression  is  greater  if the  shock  is
oblique. The  shock is  clearly oblique  for $z \not=  0$ since  it is
propagating  along  the  density  gradient,  while the  field  in  the
equilibrium core is very nearly parallel to the $z$ direction.

If $\tau_c$ is the compression in this shock, then (\ref{exactp}) gives

\begin{equation}
p = \frac{B_c^2}{2} (\tau_c^2 + \tau_c \beta_c) = p_i = M_a^2 B_i^2,
\label{pcore}
\end{equation}

\noindent
where and $B_c$ and  $\beta_c$ are the magnetic field  and $\beta$ in the
core. Solving this for $\tau_c$ gives

\begin{equation}
\tau_c = \frac{1}{2} \left[ {-\beta_c + \surd \left( {\beta_c^2 + 8
 M_a^2 \frac{B_i^2}{B_c^2} } \right) } \right]. 
\label{taucore}
\end{equation}

\noindent
For low-temperature cores close to criticality, $B_c \simeq B_i$, but
although $\beta_c < 1$, it is not really small. Nevertheless, we shall
neglect it to get

\begin{equation}
\tau_c = \surd 2 M_a.
\label{apptaucore}
\end{equation}

Since the shock is actually oblique, the post-shock velocity along the
field lines is of the order of the shock velocity except near $z = 0$.
The strength of the shock is considerably increased by convergence, so
that we  can assume  that it  is strong, in  which case  its velocity,
$V_s$, is given by

\begin{equation}
V_s^2 = \frac{p_i}{\rho_c (1 - 1/\tau_c)}.
\label{shockvel}
\end{equation}

\noindent
which becomes

\begin{equation}
V_s^2  \simeq \frac{M_a^2 B_i^2}{\rho_c  (1 -  1/\surd  2 M_a)},
\label{appshockvel}
\end{equation}

\noindent
upon  substitution for  $p_i$ and  $\tau_c$ from  (\ref{approxp}) and
(\ref{apptaucore}).

The orientation of the shock relative  to the field lines is such that
the flow  parallel to  the field  is directed towards  $z =  0$, which
means that we have two  streams with velocity $\simeq V_s$ and density
$\tau_c \rho_c$ that collide at $z = 0$. This produces a density of

\begin{equation}
\rho_{max} \simeq \tau_c \rho_c \frac{V_s^2}{c_c^2}.
\label{rhomax1}
\end{equation}

\noindent
Substituting for $\tau_c$ and $V_s^2$ from (\ref{apptaucore}) gives

\begin{equation}
\rho_{max} \simeq  2 \surd 2 \frac{M_a^3}{\beta_c(1 -  1/\surd 2 M_a)}.
\rho_c.
\label{rhomax3}
\end{equation}

\begin{figure}
\centering
\includegraphics[width=1.1\columnwidth]{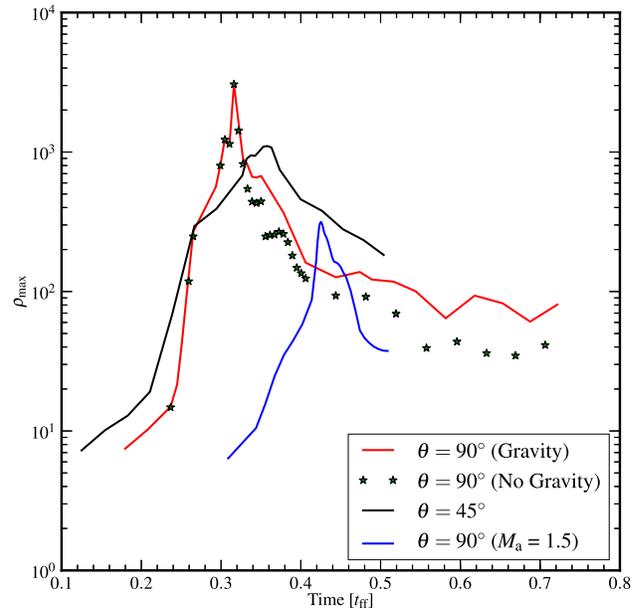}
\caption{The  maximum   density  as  a   function  of  time.}
\label{fig:maxdense}
\end{figure}

In our  case we have  $M_a = 2$,  $\rho_c = 2.08$, $\beta_c  = 0.426$,
which   gives   $\rho_{max}   =    171$.    As   we   can   see   from
Fig.~\ref{fig:maxdense}, the maximum density  is much higher than this
($\simeq 3~10^3$), which is  presumably because we have neglected both
the  effect  of shock  convergence  and  the  reflected shock  on  the
strength of the shock in the  core. 
For $M_a = 1.5$, the simulation gives $\rho_{max}$ = 315 (see fig.~\ref{fig:maxdense}) , 
whereas the multiplication of the $M_a = 2$ simulation result for $\rho_{max}$ 
by ${\zeta^3}/(1 - 1/\sqrt 2 \zeta)$, where $\zeta = 1.5/2$, gives
$\rho_{max} = 1577$. This
disagreement  is not  too surprising  since the  assumption  of strong
shocks is  not valid for  such low Mach  numbers. It would be  nice to
look at larger values of $M_a$,  but it then becomes very difficult to
resolve the thickness  of the high density region.   All this tells us
that  (\ref{rhomax3}) only  gives a  rough indication  of  the maximum
density,  but we  have  established that  this  mechanism can  produce
surprisingly high densities even for relatively weak shocks.

Fig.~\ref{fig:maxdense}  also shows  that the  oblique  shock produces
densities of  the same order  as the perpendicular shock,  which means
that the effect  is not dependent on a precise  alignment of the shock
normal  with the  $z =  0$ plane.   Although the  oblate shape  of the
equilibrium core means that shock convergence is less important if the
shock normal is not perpendicular to the field, this is compensated by
the fact that  the density is higher behind the  more oblique shock in
the core.

In Fig.~\ref{fig:maxdense}  we have  also plotted the  maximum density
for  the perpendicular case  with self-gravity  switched off  once the
shock  begins to  interact with  the  core.  This  clearly shows  that
self-gravity has no  effect on the evolution up to  the point at which
the  maximum  density is  reached,  but that  it  does  slow down  the
subsequent re-expansion. In both cases  the very high density does not
persist for long,  but the density is still  substantially larger than
the initial value even at the latest times.

\section{Discussion and Conclusions}
\label{sec:discuss}

The main result  of this paper is that,  even for weak shocks,
shock focussing  leads to surprisingly  large increases in  density in
shock-core  interactions.   This  is   entirely  an  MHD  effect,  but
self-gravity is  nevertheless essential to  the process. The  shock is
focussed by  the density gradient  in the gravitationally  bound cloud
and  the  re-expansion  of  the  dense  region  is  prevented  by  its
self-gravity.

\cite{Chen:2012p11053}  have  argued  that  ambipolar diffusion  in  a
time-dependent  shock can  lead to  a transition  from  a magnetically
sub-critical to  magnetically super-critical state.   The inclusion of
ambipolar  diffusion  and  Hall   processes  in  future  work  of  the
interactions of  shocks with cores  will therefore be  of considerable
interest.   The very  large increases  in density  that we  have found
suggest  that the  results of  \cite{Chen:2012p11053},  who considered
plane-parallel  flows only, provide  rather conservative  estimates of
the extent to which transient effects in shocks are likely to increase
the mass-to-flux ratio.

\bibliographystyle{mn2e}
%\bibliography{/Users/bhargavvaidya/MyProject/work/Leeds_Uni/Science/Sam_Data/Paper/ref_PaperI}

\begin{thebibliography}{}

\bibitem[\protect\citeauthoryear{{Ashmore}, {Van Loo}, {Caselli}, {Falle} \&
  {Hartquist}}{{Ashmore} et~al.}{2010}]{Ashmore:2010p7041}
{Ashmore} I.,  {Van Loo} S.,  {Caselli} P.,  {Falle} S.~A.~E.~G.,
  {Hartquist} T.~W.,  2010, \aap, 511, A41

\bibitem[\protect\citeauthoryear{Birkhoff, MacDougall, Pugh \& Taylor}{Birkhoff
  et~al.}{1948}]{Birkhoff}
Birkhoff G.,  MacDougall D.,  Pugh E.,    Taylor G.,  1948, Journal of Applied
  Physics, 19, 563

\bibitem[\protect\citeauthoryear{{Chapman}, {Goldsmith}, {Pineda}, {Clemens},
  {Li} \& {Kr{\v c}o}}{{Chapman} et~al.}{2011}]{Chapman:2011p11866}
{Chapman} N.~L.,  {Goldsmith} P.~F.,  {Pineda} J.~L.,  {Clemens} D.~P.,  {Li}
  D.,    {Kr{\v c}o} M.,  2011, \apj, 741, 21

\bibitem[\protect\citeauthoryear{{Chen} \& {Ostriker}}{{Chen} \&
  {Ostriker}}{2012}]{Chen:2012p11053}
{Chen} C.-Y.,  {Ostriker} E.~C.,  2012, \apj, 744, 124

\bibitem[\protect\citeauthoryear{{Dedner}, {Kemm}, {Kr{\"o}ner}, {Munz},
  {Schnitzer} \& {Wesenberg}}{{Dedner} et~al.}{2002}]{Dedner:2002p9747}
{Dedner} A.,  {Kemm} F.,  {Kr{\"o}ner} D.,  {Munz} C.-D.,  {Schnitzer} T.,
  {Wesenberg} M.,  2002, Journal of Computational Physics, 175, 645

\bibitem[\protect\citeauthoryear{{Falle}, {Hubber}, {Goodwin} \&
  {Boley}}{{Falle} et~al.}{2012}]{Falle:2012p12513}
{Falle} S.,  {Hubber} D.,  {Goodwin} S.,    {Boley} A.,  2012, in {Pogorelov}
  N.~V.,  {Font} J.~A.,  {Audit} E.,   {Zank} G.~P.,  eds, Numerical Modeling
  of Space Plasma Slows (ASTRONUM 2011) Vol.~459 of Astronomical Society of the
  Pacific Conference Series, {Comparison Between AMR and SPH}.
p.~298

\bibitem[\protect\citeauthoryear{{Falle}, {Komissarov} \& {Joarder}}{{Falle}
  et~al.}{1998}]{Falle:1998p4496}
{Falle} S.~A.~E.~G.,  {Komissarov} S.~S.,    {Joarder} P.,  1998, \mnras, 297,
  265

\bibitem[\protect\citeauthoryear{{Fragile}, {Anninos}, {Gustafson} \&
  {Murray}}{{Fragile} et~al.}{2005}]{Fragile:2005p10978}
{Fragile} P.~C.,  {Anninos} P.,  {Gustafson} K.,    {Murray} S.~D.,  2005,
  \apj, 619, 327

\bibitem[\protect\citeauthoryear{{Gregori}, {Miniati}, {Ryu} \&
  {Jones}}{{Gregori} et~al.}{2000}]{Gregori:2000p10754}
{Gregori} G.,  {Miniati} F.,  {Ryu} D.,    {Jones} T.~W.,  2000, \apj, 543, 775

\bibitem[\protect\citeauthoryear{{Kwak}, {Shelton} \& {Raley}}{{Kwak}
  et~al.}{2009}]{Kwak:2009p10827}
{Kwak} K.,  {Shelton} R.~L.,    {Raley} E.~A.,  2009, \apj, 699, 1775

\bibitem[\protect\citeauthoryear{{Lim}, {Falle} \& {Hartquist}}{{Lim}
  et~al.}{2005}]{Lim:2005p5897}
{Lim} A.~J.,  {Falle} S.~A.~E.~G.,    {Hartquist} T.~W.,  2005, \apjl, 632, L91

\bibitem[\protect\citeauthoryear{{Marchwinski}, {Pavel} \&
  {Clemens}}{{Marchwinski} et~al.}{2012}]{Marchwinski:2012p11767}
{Marchwinski} R.~C.,  {Pavel} M.~D.,    {Clemens} D.~P.,  2012, \apj, 755, 130

\bibitem[\protect\citeauthoryear{{Mouschovias}}{{Mouschovias}}{1976a}]{Mouscho%
vias:1976p8743}
{Mouschovias} T.~C.,  1976a, \apj, 206, 753

\bibitem[\protect\citeauthoryear{{Mouschovias}}{{Mouschovias}}{1976b}]{Mouscho%
vias:1976p8742}
{Mouschovias} T.~C.,  1976b, \apj, 207, 141

\bibitem[\protect\citeauthoryear{{Mouschovias} \& {Spitzer} Jr.}{{Mouschovias}
  \& {Spitzer}}{1976}]{Mouschovias:1976p9250}
{Mouschovias} T.~C.,  {Spitzer} Jr. L.,  1976, \apj, 210, 326

\bibitem[\protect\citeauthoryear{{Shelton}, {Kwak} \& {Henley}}{{Shelton}
  et~al.}{2012}]{Shelton:2012p10934}
{Shelton} R.~L.,  {Kwak} K.,    {Henley} D.~B.,  2012, \apj, 751, 120

\bibitem[\protect\citeauthoryear{{Shin}, {Stone} \& {Snyder}}{{Shin}
  et~al.}{2008}]{Shin:2008p9750}
{Shin} M.-S.,  {Stone} J.~M.,    {Snyder} G.~F.,  2008, \apj, 680, 336

\bibitem[\protect\citeauthoryear{{Van Loo}, {Falle} \& {Hartquist}}{{Van Loo}
  et~al.}{2010}]{VanLoo:2010p9828}
{Van Loo} S.,  {Falle} S.~A.~E.~G.,    {Hartquist} T.~W.,  2010, \mnras, 406,
  1260

\bibitem[\protect\citeauthoryear{{Van Loo}, {Falle}, {Hartquist} \&
  {Moore}}{{Van Loo} et~al.}{2007}]{VanLoo:2007p9421}
{Van Loo} S.,  {Falle} S.~A.~E.~G.,  {Hartquist} T.~W.,    {Moore} T.~J.~T.,
  2007, \aap, 471, 213

\bibitem[\protect\citeauthoryear{{Yu}, {Lou}, {Bian} \& {Wu}}{{Yu}
  et~al.}{2006}]{Yu:2006p12405}
{Yu} C.,  {Lou} Y.-Q.,  {Bian} F.-Y.,    {Wu} Y.,  2006, \mnras, 370, 121

\end{thebibliography}

\label{lastpage}
%---------------------------------------------------------------

\end{document}